\documentclass[12pt,preprint]{aastex}
\newcommand\etal{{\it et al.}}
\newcommand\lae{\mathrel{<\kern-1.0em\lower0.9ex\hbox{$\sim$}}}
\newcommand\gae{\mathrel{>\kern-1.0em\lower0.9ex\hbox{$\sim$}}}
\newcommand\kms{km~s$^{-1}$}
\newcommand\mone{$^{-1}$}
\newcommand\mtwo{$^{-2}$}

\newcommand{\Lya}{Ly$\alpha$}
\newcommand{\Ha}{H$\alpha$}
\newcommand{\Hb}{H$\beta$}

\newcommand{\subsun}{$_\odot$}

\begin{document}

\title{ HST/STIS Spectroscopy of the \Lya\ Emission Line in the 
Central Dominant Galaxies in A426, A1795, and A2597: Constraints
on Clouds in the Intracluster Medium\footnote{Based on observations make with 
the NASA/ESA Hubble Space Telescope, obtained at the Space Telescope Science
Institute which is operated by the Association of Universities for Research
in Astronomy, Inc, under NASA contract NAS 5-26555. These observations are
associated with program \#8107.}
}
\author{Stefi A. Baum}
\affil{ Center for Imaging Science, Rochester Institute of Technology }

\author{Ari Laor }
\affil{Technion-Israel Institute of Technology}
\authoraddr{Department of Physics, Haifa, 32000, Israel}

\author{Christopher P. O'Dea}
\affil{ Department of Physics, Rochester Institute of Technology }

\author{Jennifer Mack \& Anton M. Koekemoer}
\affil{ Space Telescope Science Institute\footnote{Operated
by the Association of Universities for Research in
Astronomy, Inc. under contract NAS 5-26555 with the National Aeronautics and
Space Administration.}}


\begin{abstract}
We report on HST/STIS spectra of the \Lya\  emission in the central dominant 
galaxies in three rich clusters of galaxies. 
We find evidence for a population of clouds in the intracluster medium.
We detect 10 \Lya\ absorption systems towards the nucleus of NGC1275 with 
columns of N(HI) $\sim 10^{12}$ to $10^{14}$ cm\mtwo. These columns would
not have been detected in the 21 cm line, but are easily detected in
the \Lya\ line. Most of the absorption features are 
located in the broad wings of the emission line. The detected absorption features 
are most consistent with associated nuclear absorption systems. 
There is very little nuclear absorption at the systemic velocity in NGC1275 
(Feature 8 contains N(HI) $\sim 3\times  10^{12}$  cm\mtwo). This 
implies that the large columns detected in the 21 cm line towards
the parsec scale radio source avoid the line of sight to the nucleus. 
This gas may be located in a circumnuclear disk or torus.
We detect at least one and possibly two absorption features towards 
the extended \Lya\ in A426. We do not detect \Lya\ absorption 
towards the extended \Lya\  emission in A1795, and A2597 
with upper limits N(HI) $\sim  10^{13}$  cm\mtwo\ for optically 
thin absorbers. 
Our data constrain the covering factor of any high column density gas 
(N(HI) $> 10^{15}$  cm\mtwo) in the ICM to be less than 25\%. 
Our results suggest that the lack of observed intermediate temperature gas is not 
explained by obscuration. 
In addition, the low columns of gas on the $\sim 100$ kpc scales
in the ICM suggests that (1) the rate at which cold gas accumulates in the
ICM on these scales is very low, and (2) the dense nebulae in the central
$\sim 10$ kpc must have cooled or been deposited {\it in situ}. 

\end{abstract}

\section{INTRODUCTION}

The hot T$\sim 10^{7-8}$~K X-ray emitting gas is currently thought to
constitute the bulk of the baryonic mass in rich clusters of galaxies.
If substantial amounts of cold gas do exist in the ICM, this would have
important implications for our understanding of (1) cooling flows (how much
mass do they deposit?), (2) the physics of the ICM (is it a multi-phase 
medium like our ISM?), (3) the intracluster magnetic
field (can it suppress conduction?), (4) galaxy formation and evolution
(how important are winds and stripping?), and (5) Ly${\alpha}$ forest systems
(are they formed in clusters? - e.g., Crawford \etal\ 1987).

There are several potential candidates for a population of cold
clouds which might exist in the intracluster medium (ICM) of clusters of
galaxies (e.g.,  Sarazin 1988):
(1) Cold gas which has been removed from the individual galaxies, possibly
by ram pressure stripping or by galaxy collisions (Soker, Bregman \& Sarazin 1991; 
Sparks \etal\ 1989);  (2) Primordial clouds or
protogalaxies which are currently
falling into the cluster; (3) clouds which condense from thermal
instabilities in a cooling flow in the inner $\sim 100$ kpc of
the cluster center (e.g., Cowie and Binney 1977; Fabian and Nulsen 1977;
Mathews and Bregman 1978; Fabian 1994). Mass accretion rates in cooling
flows were estimated to be in the range \.m $\sim 10 -
100 $ M\subsun\  per year.
If these accretion rates last for the lifetime of the cluster ($\sim
10^{10}$ yr) the accumulated mass in gas would be $10^{11} - 10^{12}$ 
M\subsun.
On the other hand, potential sinks and destructive processes for cold gas
include: (1) star formation (e.g., Allen 1995), (2) heating and
evaporation via thermal conduction, mixing layers, etc. (e.g., Sparks 1992;
B\"ohringer \& Fabian 1989), and shredding (e.g., Loewenstein \& Fabian 1990).

X-ray spectroscopy with XMM-Newton and Chandra has failed to find evidence
for gas at temperatures below about one-third of the cluster virial
temperature (e.g., Kaastra \etal\ 2001; Tamura \etal\ 2001; Peterson \etal\ 
2001,2003). The limits on the luminosity of the intermediate temperature
gas imply reductions in the inferred mass accretion rates by factors of 5-10.
One possibility is that there is a source of heat in the ICM
(radio galaxies? - e.g., Baum \& O'Dea 1991; Tucker \& David 1997;
Soker, Blanton \& Sarazin 2002; B\"ohringer \etal\
2002) which halts the cooling of the gas. Alternately, it may be possible to 
absorb  the X-ray luminosity of the cooling gas, 
with the luminosity finally emerging at another waveband (e.g., Fabian 
\etal\ 2001,2002; Peterson \etal\ 2001). 
This raises the question of whether the required absorbing
matter exists in these clusters. 

Early X-ray observations indicated evidence
for low energy X-ray absorption in excess of that which would be
produced by the known Galactic column density of HI (e.g., White \etal\ 1991;
Mushotzky 1992; Allen \etal\ 1993; Allen \& Fabian 1994).
White \etal\ (1991) suggested that this was evidence for  the 
existence of a population of cold 
clouds with column density $N_H \sim 10^{21} - 10^{22} $ cm$^{-2}$ 
and a covering factor of order unity. 
However, recent Chandra and XMM observations with greater sensitivity and
spectral  resolution  have not confirmed the need for excess absorption.
Current upper limits on the column density
of Hydrogen in the clusters are typically a few $\times 10^{20}$ cm$^{-2}$ 
(e.g., Kaastra \etal\ 2001, 2004; Tamura \etal\ 2001;
Peterson \etal\ 2001, 2003; Blanton, Sarazin \& McNamara 2003;  
but see Ettori \etal\ 2002 for possible intrinsic absorption in A1795).

After a decade of non-detections
molecular gas has finally been detected in clusters other than A426 (Edge 2001).
Roughly half the extreme cooling flow clusters surveyed by Edge 
exhibit CO corresponding to masses of molecular hydrogen in the range $\sim
10^9 - 10^{10}$ M$_\odot$. 
Intensive searches for the 21 cm line of Atomic hydrogen in both emission and
absorption have placed limits on HI columns of $\lae 10^{19}$ cm\mtwo\
(e.g., Burns, White, and Haynes  1981; Valentijn
and Giovanelli 1982; McNamara, Bregman, and O'Connell 1990; Jaffe 
1991, 1992;  Dwarakanath \etal\ 1994; O'Dea, Gallimore, \& 
Baum 1995; O'Dea, Payne \& Kocevski 1998).  HI absorption has been detected
in only a few clusters (A426/3C84 - Crane \etal\ 1982; Jaffe 1990; 
Sijbring 1993; A780/Hydra A - Taylor 1996; A2597/PKS2322-123 - O'Dea, 
Baum \& Gallimore 1994; Taylor \etal\ 1999) 
with implied masses of atomic hydrogen in the range $\sim 10^7 - 10^8$  
M$_\odot$. The strong dominance of molecular relative to atomic gas in these 
clusters is in contrast
to the situation in normal galaxies where roughly $\sim 15\%$ of the cold gas
is thought to be in molecular form (e.g., Boselli, Lequeux and Gavazzi 2002); however,
the bulk of the molecular gas in galaxies may be undetected
(e.g., Allen \etal\ 1997).


The 21 cm limits on the HI column are subject to two caveats: 1) the
limits depend linearly on the electron excitation temperature, which
may be higher than previously thought. 2) the absorbing gas may have
a very small velocity dispersion, producing a very narrow saturated
absorption line, and thus a lower absorption equivalent width
than estimated for the optically thin case.
The UV region provides  a powerful but as  yet not fully exploited probe of
the ICM. The \Lya\ absorption cross-section is $\gae 10^7$ times larger
than that of the 21 cm line (e.g., Bahcall \& Ekers 1969; Laor 1997).
In addition, it is not subject to the two caveats which affect the 21~cm
line, as 1) the absorption is practically independent of the gas temperature
(the extremely short lifetime of the excited $n=2$ level ensures that
effectively all HI is in the ground level); 2) significant absorption is
expected even if the absorber has no velocity dispersion (the
absorption will occur in the Lorentzian, or ``damping'' wings). Thus,
 \Lya\ is superior to the 21 cm line as a probe of the cold gas content of
the ICM. 

In fact, Koekemoer \etal\ (1998) presented HST/FOS UV spectra of the quasar in
the center of A1030 and placed limits on column densities in the range
$10^{11} - 10^{13}$ cm\mtwo\ for a wide range of molecular, atomic, and ionized 
species that may be associated with the ICM.
Miller, Bregman, \& Knezek (2001) obtained  limits of $10^{12} - 10^{13}$ cm\mtwo\ 
on column densities for several absorption lines (FeII, MgII) through
lines of sight towards the outer parts of six clusters. 
Laor (1997) has used a low resolution ($\sim 250$ \kms ) {\em HST}/FOS
spectrum of the \Lya\ line in the center of the Perseus cluster
(published by Johnstone \& Fabian 1995) to set an upper  limit of
$N_{HI} \lae 4\times 10^{17}$ cm\mtwo. It was clear that an improvement of
several orders of magnitude in the constraints on the atomic hydrogen columns
could be achieved by obtaining higher resolution spectra of the \Lya\ line. 

In this paper we present medium dispersion (20 \kms) HST/STIS long-slit
spectroscopy of the \Lya\ line towards the central nebulae in three
clusters - A426, A1795, and A2597. We report the discovery of   
\Lya\ absorption systems towards the nucleus and extended \Lya\ emission
of NGC1275. We discuss the nature and origin of these systems. We also 
place constraints on the properties of clouds in the ICM. 


\section{OBSERVATIONS AND RESULTS}

The properties of the three central galaxies are summarized in 
Table~\ref{tabsource}.
The HST/STIS (Kimble \etal\ 1998) 
spectroscopic observations are summarized in Table~\ref{tabobs}. 
We observed with the Far-UV MAMA and the G140M grating centered on the 
redshifted
\Lya\ emission line. Spectral and spatial pixel scales are 
0.05\AA\ per pixel and 0\farcs 029 per pixel, respectively. Spectral 
resolution is
1.6 pixels (0.08\AA) FWHM which gives approximately  20 \kms. 
We used the 0\farcs1 slit to preserve our spectral and spatial
resolution.
 We observed in TIME-TAG mode to allow us to reject data with 
high sky background. 

We obtained ``early-acquisition" images of the \Lya\ emission line and the 
nearby Far-UV continuum. These data are discussed by O'Dea \etal\ (2004). 
We used these images to select a position angle for the slit which maximized 
the amount of bright \Lya\ and/or FUV continuum in the slit.  
The data were reduced with the STIS pipeline using the best updated 
reference files.  

We detect spatially extended \Lya\ emission in our spectra of all
three sources. The fluxes for the extended emission are consistent with those
derived from those locations in the UV images (O'Dea \etal\ 2004).
In the A2597 spectrum, we also detect the geocoronal OI lines at 1302, 
1305, and 1306 \AA\  (Table~\ref{tabgeo}). 
In A426 we detect a bright  \Lya\ emission line with broad wings 
from the active nucleus. We do not see evidence for
point-like nuclear emission in A1795 or A2597.


\section{ABSORPTION LINES TOWARDS THE NUCLEUS OF NGC1275}

\begin{figure}
\rotate
\plottwo{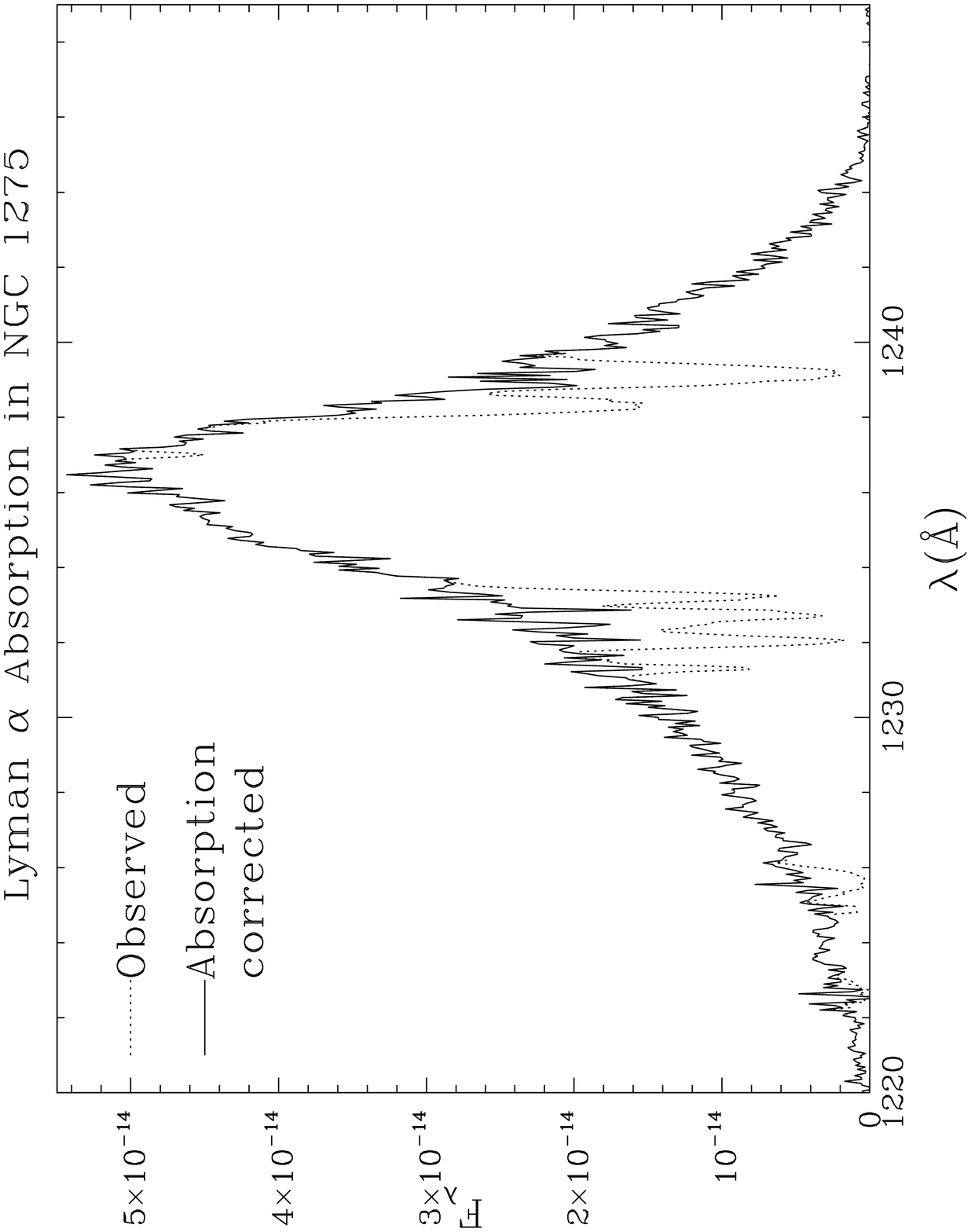}{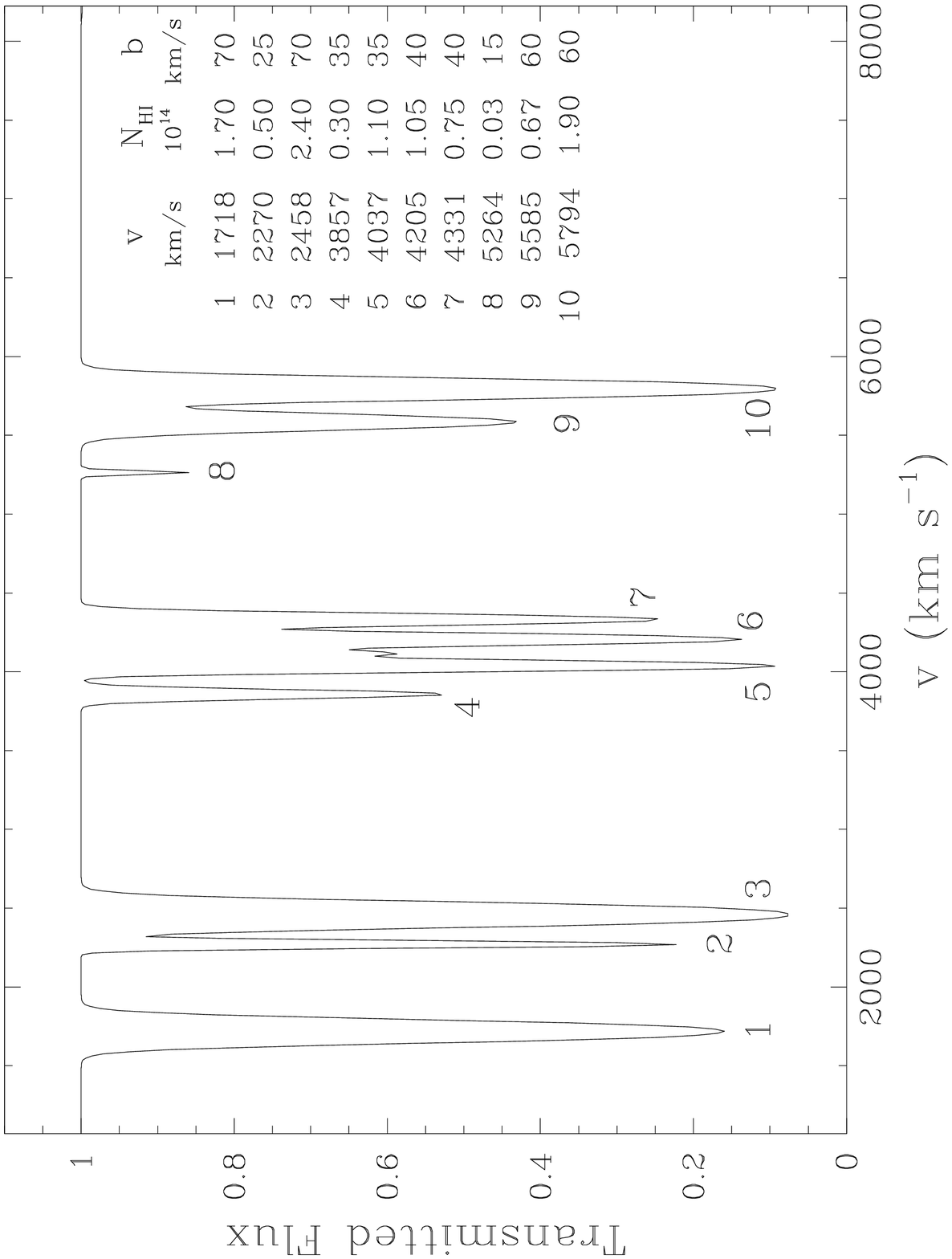}
\caption{A426. (Left). The observed and absorption corrected \Lya\ profile
of NGC1275. The spectrum is integrated in the cross-slit direction since the 
light distribution is consistent with a point source emission. 
(Right). The fraction of transmitted flux vs. observer frame velocity. The 
table on the right lists the parameters of the 10 absorption systems 
detected, including the central velocity, total HI column, and Doppler
b parameter. Our 3$\sigma$ limit on column density is $1\times 10^{12}$
cm\mtwo.
}
\label{abs}
\end{figure}

We detect ten \Lya\ absorption systems towards the nucleus 
of NGC1275 (3C84)  (Figure \ref{abs} and Table \ref{tablines}). 
In order to determine the properties of each absorption system
we multiplied the observed spectrum by $e^\tau_\lambda$, where
$\tau_\lambda$ is the wavelength dependence of the optical depth of
each system. The value of $\tau_\lambda$ is set by center of
the line, $\lambda_0$, the velocity dispersion parameter, and
the H~I column, through the standard Voigt profile calculations
(e.g., Rybicki \& Lightman 1979). 
We iterate over the values of these three parameters
until the corrected spectrum appears featureless. We estimate the
uncertainty in the column density and velocity dispersion to be at
a level of ~10-30\%. The wavelength uncertainty is generally set by 
the spectral resolution. This absorption
correction procedure assumes absorption by a uniform screen. The
implied column could be larger in case of partial obscuration
(i.e., absorbing filaments which are smaller than the continuum
source). The requirement that the \Lya\  damping wings are 
not broader than the observed absorption width provides a strict
upper limit to the possible column. The two strongly damped systems  
correspond to Galactic absorption, and to absorption by the
infalling foreground system at 8200~km/s, also detected in H~I.
The column densities in the detected absorption lines 
are in the range $N(HI) \sim 10^{12}$ 
to $10^{14}$ cm\mtwo\ and Doppler parameters are in the range
$b \simeq 25-70$ \kms. 

\begin{figure}
\plottwo{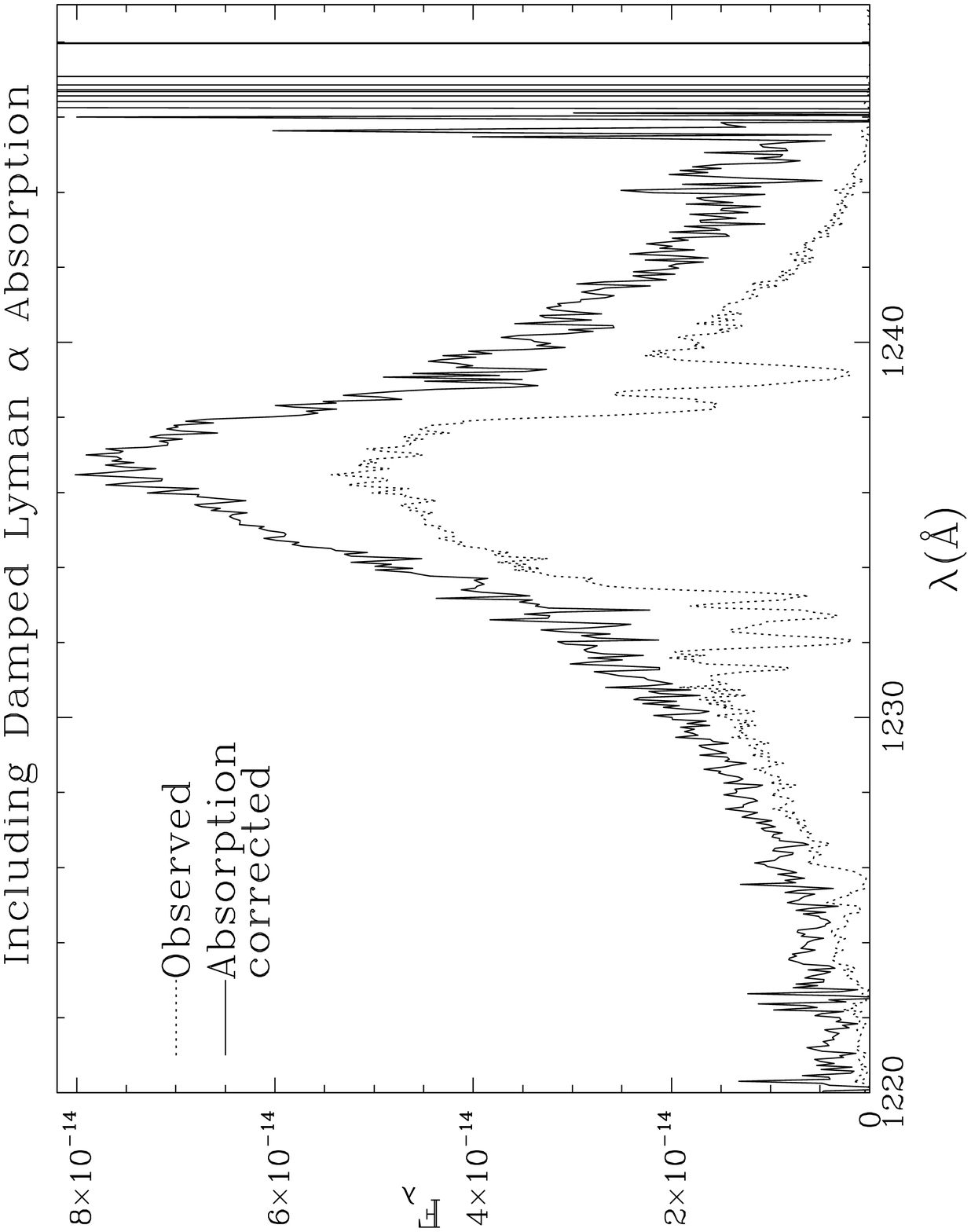}{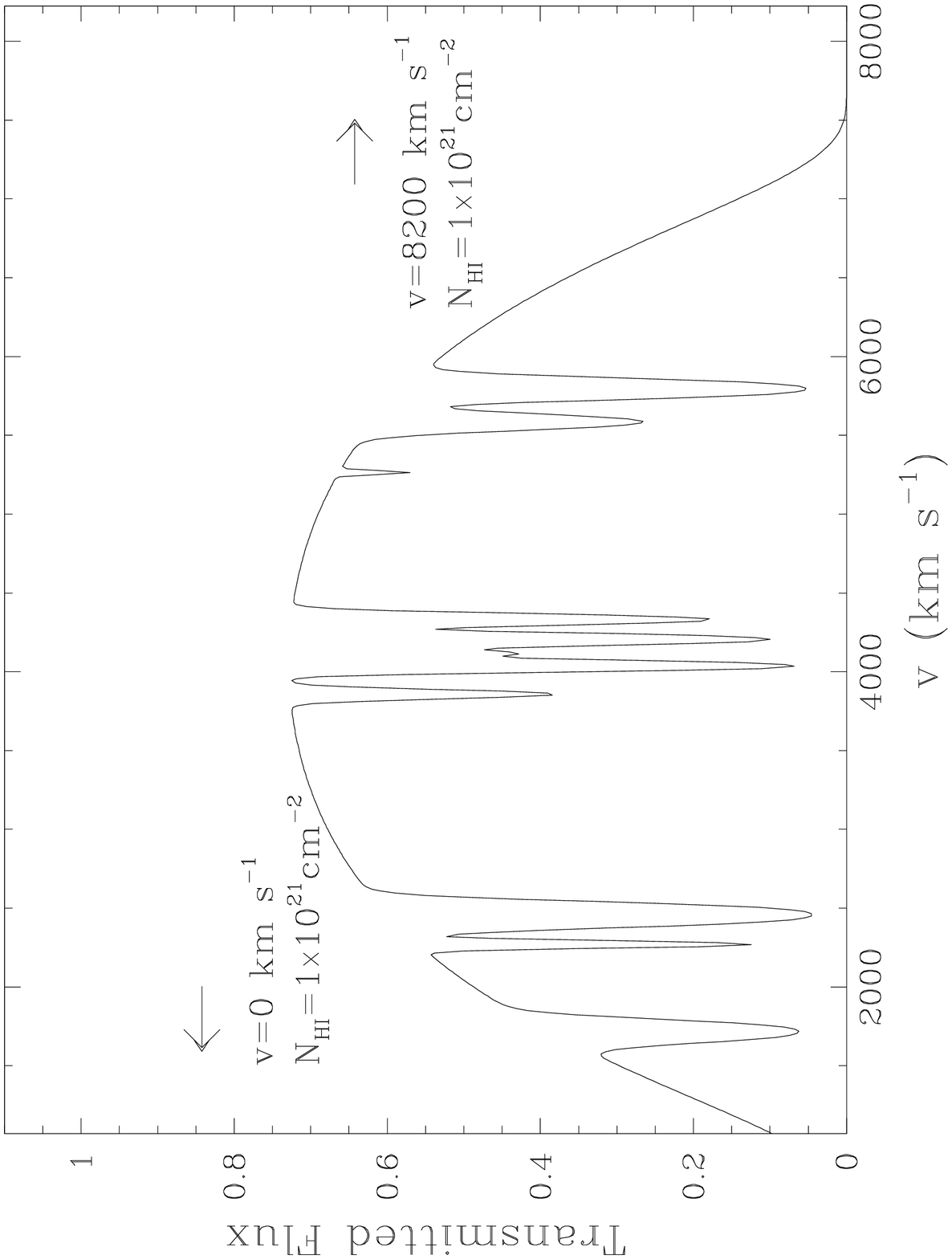}
\caption{A426.  As in Figure \ref{abs} but including corrections 
for damped \Lya\ absorption by the Galaxy (at v=0), and by the 
foreground infalling galaxy (at v=8200 \kms). The  \Lya\ absorption 
profile of both systems suggests a column of about $1\times 10^{21}$
cm\mtwo,  consistent with the 21 cm absorption by these systems.
}
\label{abs1}
\end{figure}

The \Lya\ in NGC1275 is viewed through two damped \Lya\ absorption
systems - one in our Galaxy at v=0 and one in the infalling
high velocity system (v=8200 \kms). The data corrected for
the effects of these two systems is shown in Figure \ref{abs1}.
We estimate column densities of $\sim 1 \times 10^{21}$ cm\mtwo\
in the damped systems which is in agreement with the observed 
21 cm columns, providing
a consistency check on our analysis.  

We detect \Lya\ absorption (feature 8) at the systemic velocity of 
NGC1275. However, it is the weakest of our detections. The other 
absorption features are seen mainly in the wings of the \Lya\ 
emission line. 
Seven of nine features are blue shifted with respect to the 
systemic velocity of NGC1275. The absorption systems extend out 
to -3500 \kms\ with respect to systemic.  

\subsection{Where are the nuclear 21 cm absorbers?}

The 21 cm line has been detected in absorption towards the nucleus
of NGC1275 (Crane \etal\ 1982; Jaffe 1990; Sijbring 1993) with an
optical depth $\tau \simeq 0.0024$ and FWHM$\sim 440$ \kms, which gives
a column density of $N(HI) \simeq  2\times 10^{20}$ cm\mtwo\ for 
``warm" clouds (T=100 K). However, the column measured in the 
\Lya\ absorption line (system 8) is only $N(HI) \simeq  
3\times 10^{12}$ cm\mtwo. This difference suggests that the atomic 
hydrogen detected in the 21 cm line avoids the line-of-sight to 
the nucleus.  NGC1275 exhibits a two-sided 
(but asymmetric) parsec scale radio source (e.g., Walker, Romney \&
Benson 1994; Vermeulen, Readhead \& Backer 1994). 
Our results imply that the 21 cm absorption is seen against the
parsec scale radio source, but not directly against the \Lya\
emitting region in the nucleus. 

Multi-frequency VLBI observations have shown that there is free-free
absorption towards the northern radio jet (e.g.,  Walker, Romney \&
Benson 1994; Vermeulen, Readhead \& Backer 1994; Walker \etal\ 2000). 
The free-free absorbing medium seems likely to be located in a disk or torus
(e.g., Levinson, Laor \& Vermeulen 1995; Walker \etal\ 2000). 
We suggest that the atomic hydrogen responsible for the 21 cm absorption 
may also be located in that putative disk or torus.

\subsection{Comparison of the STIS and FOS Spectra at Similar Resolution}

Johnstone \& Fabian (1995) presented an HST FOS spectrum of the nucleus
of NGC1275 taken 1993 February 3 through the G130H grating with a 
spectral resolution of $\sim 1$\AA. They found an apparent dip in 
the spectrum at 1237\AA\ which they attributed to the presence of 
a blue shifted component of \Lya\ produced by Fermi-accelerated 
\Lya\ emission (Neufeld \&
McKee 1988; Binette, Joguet \& Wang 1998). For NGC1275, this model
requires that the \Lya\  photons be scattered back and forth across
a shock with velocity $v_s \sim 100$ \kms, and atomic hydrogen column
density $1.1 \times 10^{20} < N < 1.3 \times 10^{21}$ cm\mtwo\ (Johnstone
\& Fabian 1995). The \Lya\ photons must also be generated in or near
the shock to avoid destruction by dust associated with these high
columns. 

\begin{figure}
\plotone{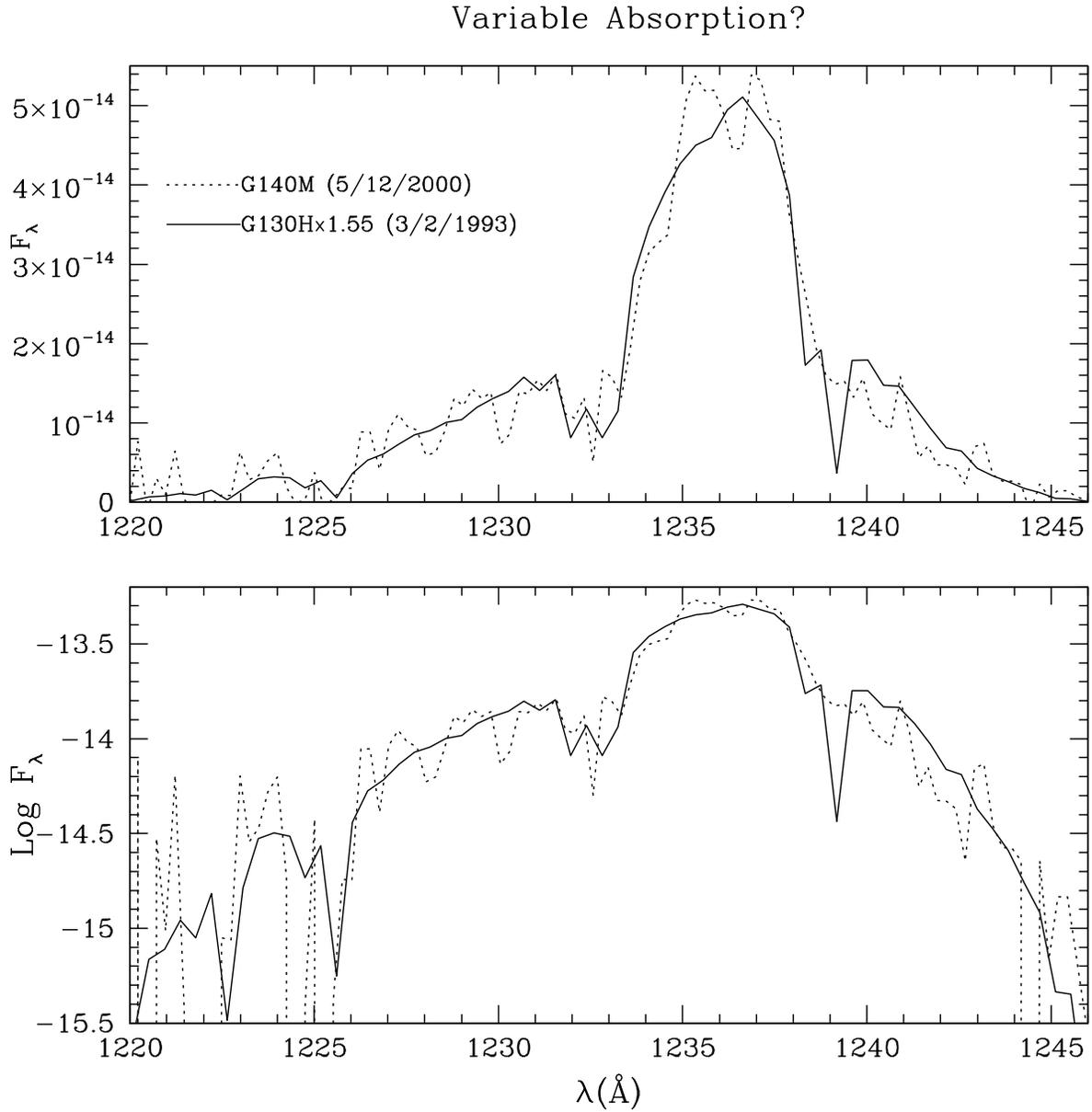}
\caption{A426. Comparison of the 1993 FOS G130H spectrum (Johnstone \&
Fabian 1995) with our 2000 STIS G140M spectrum. The FOS spectrum was scaled
by a factor of 1.55 to match the STIS spectrum.
The STIS G140M spectrum has been rebinned by a factor of eight (pixel size
0.85\AA) so that it has similar resolution to the FOS G130H spectrum.  
}
\label{var2}
\end{figure}

In Figure \ref{var2} we present a comparison of the FOS 130H data with 
our STIS G140M spectrum scales to similar spectral resolution. The FOS
spectrum is too low resolution and too low S/N to have detected the 10
\Lya\ absorption features that we detect in our STIS G140M spectrum. 
However, we should have seen the feature in the FOS spectrum at 1237\AA.
The discrepancy could be due to bad data in the FOS spectrum. However,
it is also possible that the 1237\AA\ feature is variable. This would
imply that it is an ``intrinsic absorption feature" (see \S~\ref{intrins}).
If the variability of the feature is due to the motion of clouds near
the nucleus, at a velocity of 0.01c perpendicular to our line of sight,
they could move only 0.07 light
years in the $ \sim 7$ years between the two observations. Thus, the clouds
should be less than 0.025 pc across. Alternately, the difference in the STIS
and FOS spectra could be at least partially due to aperture affects. 
The larger FOS aperture contains a contribution from extended \Lya\ emission
which exhibits a \Lya\ absorption feature at the systemic velocity (see 
below). This could be the source of the feature seen in the FOS spectrum.

\subsection{What are the Absorption Features Towards NGC1275?} 
\subsubsection{Intrinsic Nuclear Absorption Features}\label{intrins}

Seven of nine features seen towards the nucleus of NGC1275 are 
blue shifted with respect to the systemic velocity of NGC1275. 
The absorption systems extend out to -3500 \kms\ with respect 
to systemic.  
One possibility which applies to these  systems is that
they are associated with out flowing nuclear gas. 
We note that ``intrinsic" absorption features are found in about
half of Seyfert I AGN (e.g., Crenshaw \etal\ 1999) - which have 
similar optical and X-ray luminosity as NGC1275. The intrinsic
absorption features in AGN tend to be blue-shifted, variable,
have a range of line widths of $\sim 20-400$ \kms, and tend to 
exhibit high ionization metal lines (e.g., Weymann \etal\ 1997; 
Crenshaw \etal\ 1999; Kriss \etal\ 2000). 
The intrinsic features are thought to  be produced by clouds in an 
out-flowing wind. 

The column densities, widths, and velocity offsets of the absorption
features in NGC1275 are roughly consistent with those of intrinsic
absorption features. In addition, the possibility that the 1237\AA\
feature (if real) is variable is also consistent with that origin. 
Additional observations to search for variability
and determine the metalicity of the systems would test this scenario. 

\subsubsection{Emission Line Filaments}

We have compared the velocities of the absorption features with those
of the emission line filaments given by Conselice, Gallagher \& Wyse
(2001). The emission line filaments cover a velocity range of roughly
5000 to 5500 \kms. We find that only one absorption feature (system 8) agrees
with the velocity of any of the filaments. In fact most of the absorption
features lie at velocities well beyond the range of velocities of
the emission line filaments. Thus, the emission line filaments do not
produce the majority of the absorption features.  

The radio source 3C84 appears to have evacuated ``bubbles" in the ICM
(e.g., B\"ohringer  \etal\ 1993; McNamara, O'Connell \& Sarazin 1996;
Fabian \etal\ 2000). Fabian (private communication) has noted that 
we may be viewing the nucleus of NGC1275 through the northern
bubble and has suggested that the bubble has displaced the emission
line filaments from our line-of-sight to the nucleus. 
An alternate possibility is that any dense line-emitting clouds which 
have been engulfed by the expanding bubble have been destroyed by
shredding (e.g., Loewenstein \& Fabian 1990; Klein, McKee \& Colella 
1994; O'Dea \etal\ 2003, 2004). 

\subsubsection{Galaxies in the Perseus Cluster}

We have searched in NED for galaxies close to NGC1275 in projection
and in velocity which  might be candidates for some of the absorption
features (Table \ref{tabgalaxy}). We searched a box 8 arcmin across
centered on NGC1275 and within a heliocentric velocity range 1000 to 6000
\kms. We find that  4 of the absorption systems have potential
associations with cluster galaxies 30-80 kpc in projection to the
line-of-sight and 27-106 \kms\ offset in velocity. At this point it
is not clear which if any of these candidate associations is real. 

\subsubsection{\Lya\ Forest Systems}

The column densities and Doppler b parameters of the absorption features 
towards  NGC1275 are similar to those of typical \Lya\ forest systems
(e.g., Rauch 1998). However, 
according to Bahcall \etal\  (1996) the density of the \Lya\  forest
at $z\sim 0$ is $dN/dz=24.5 \pm 6.6$ for systems with $EW>0.25A$. 
Now, the 10 absorption systems we see cover $\Delta v = 4076$ km/s, or 
$dz = \Delta v/c=0.0136$.
Thus, we would have expected 0.0136 x 24.5=0.33 \Lya\ forest
systems to fall at the velocity interval we observed, while there
are 7 (out of the 10) systems with $EW>0.25A$. So, this suggests
that the systems we have detected are most likely
not part of the general \Lya\ forest in the direction of A426.
Since the velocity range of the \Lya\  absorption systems is so 
much larger than the range of velocities of the \Ha\ emission filaments 
(Conselice \etal\ 2001), it seems that the best bet is that most of these 
absorption systems are associated with the active nucleus.

\section{CONSTRAINTS ON EXTENDED ABSORPTION IN A426, A1795 AND A2597}

\begin{figure}
\plotone{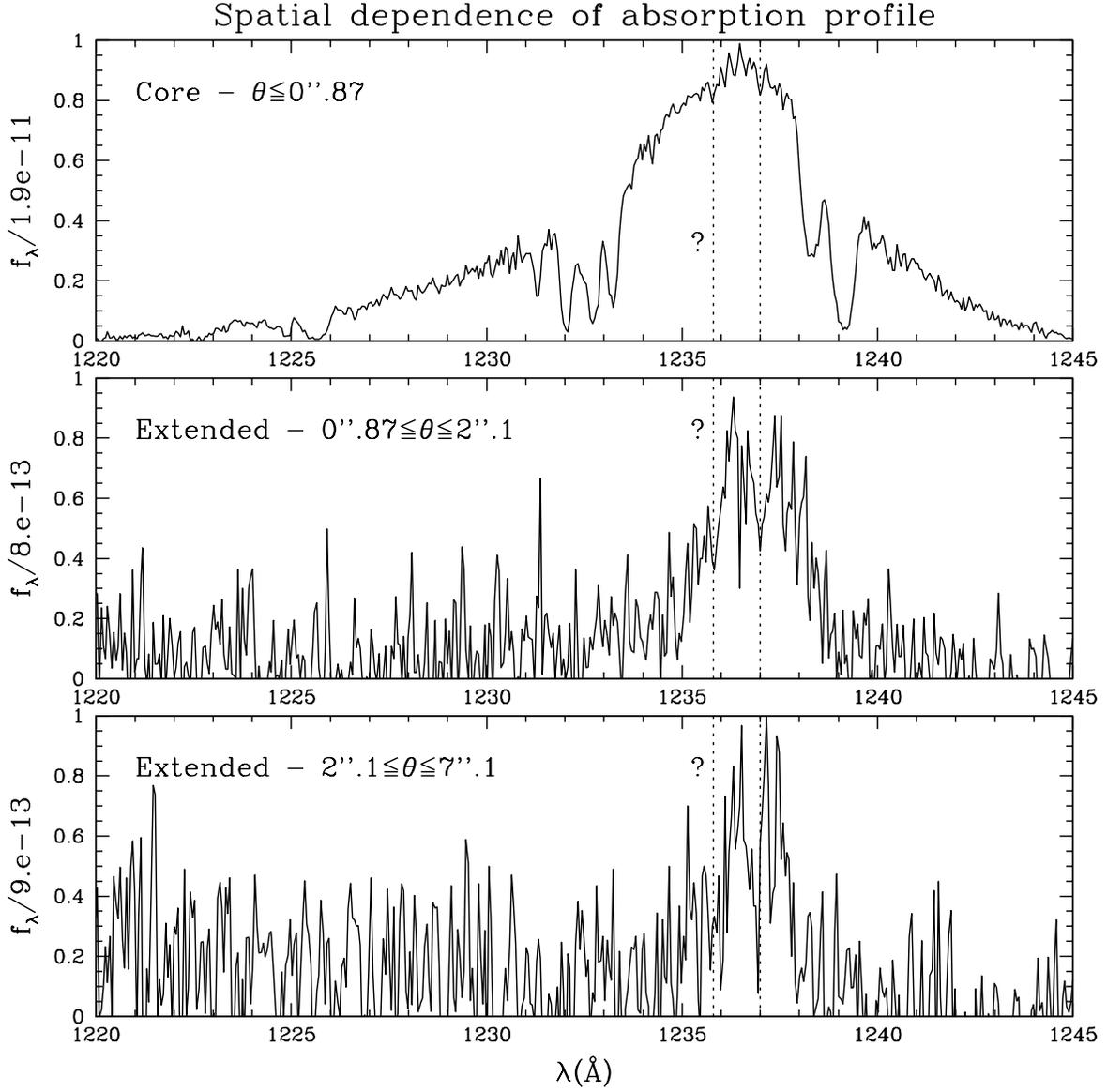}
\caption{A426 G140 spectra. (Top) Nucleus of NGC1275. 
(Middle) Sum of extended emission between 0.87 and 2.1 arcsec
from the nucleus. (Bottom) (Middle) Sum of extended emission between 
2.1 and 7.1 arcsec from the nucleus. }
\label{a426diff}
\end{figure}

Figure~\ref{a426diff} shows a comparison of the nuclear spectrum 
with the sum of two off-nuclear emission regions. We detect the 
absorption system seen against the nucleus at the systemic 
velocity of  5264 \kms\ and possibly also a system at  5259 \kms\
which is not formally detected against the nucleus. 
The column density in the extended 5264 \kms\ system is about 
a factor of 30 times larger than that seen towards the nucleus. 
This suggests that the extended absorbing system is possibly
associated with the plane of the galaxy.  If the extended system
is damped and is seen with an absorption  depth of 0.5 because
of scattering or partial covering, then the upper limit to
the estimated column becomes $N(HI) < 2 \times 10^{18}$ cm\mtwo. 

\begin{figure}
\plotone{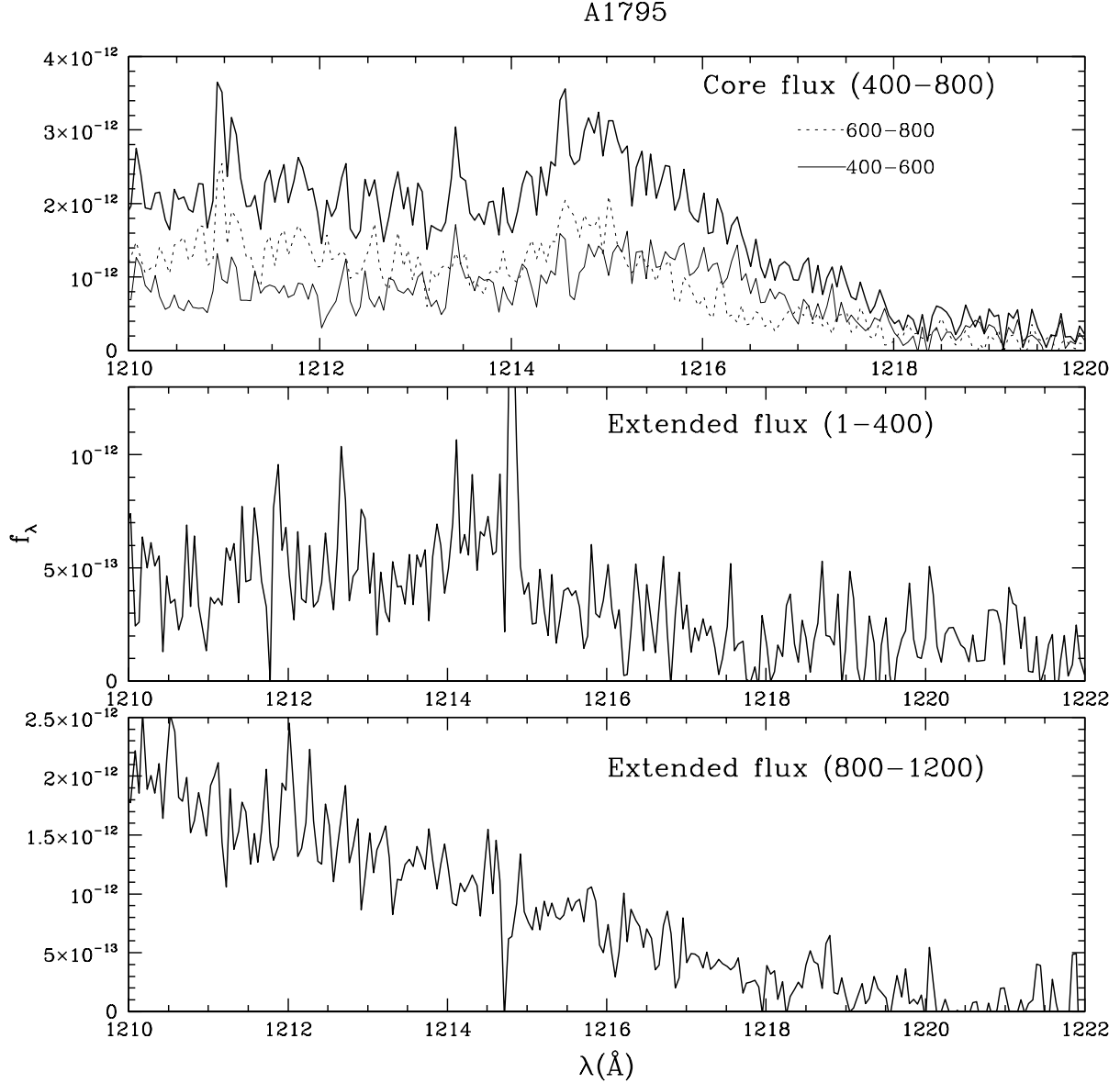}
\caption{A1795. 
The A1795 G140 spectra versus rest wavelength.
Each panel shows a spectrum integrated along one third of the 
slit, corresponding to 11.6 arcsec (row numbers are indicated 
in parenthesis, where row 1 corresponds to the bottom of the
image in Fig.~5, south). Significant \Lya\ emission is detected
near the core. The peak of the emission from the upper half
of the core (rows 600-800) is blueshifted by $\sim 200$ km/s 
compared to the emission from the lower half (rows 400-600).
Some extended emission may be present south of the nucleus
region, but none is detected to the north. The apparent rise
in the spectrum in the south results from a non uniform
instrumental background. No clear \Lya\ absorption is
present towards the nucleus.
}
\label{a1795tot}
\end{figure}

\begin{figure}
\plotone{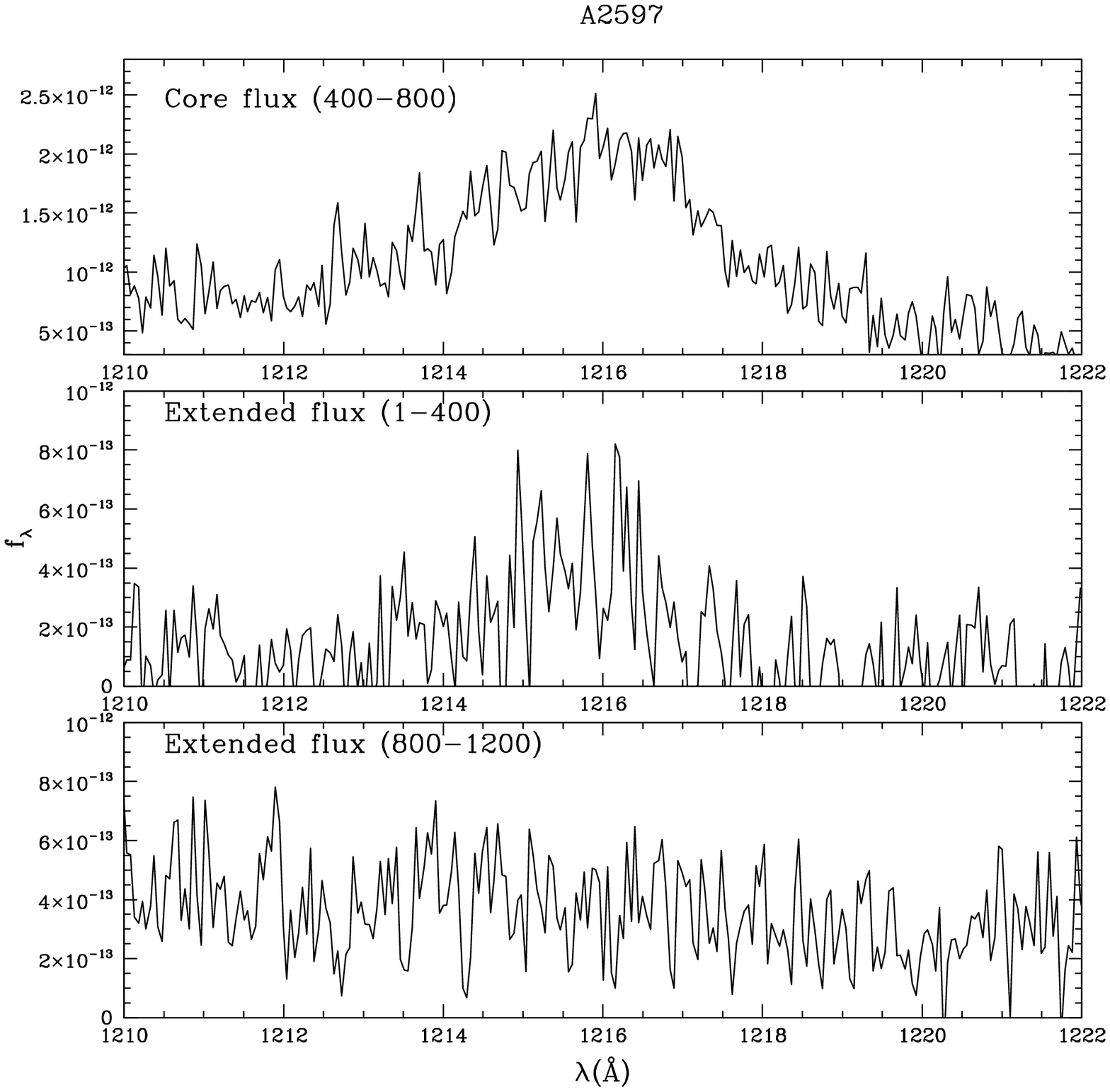}
\caption{A2597. 
As in Fig.~\ref{a1795tot} for A2597. Extended emission is present
to the south, but none to the north. No clear \Lya\  
absorption is present towards the nucleus.
}
\label{a2597tot}
\end{figure}

Extracted spectra showing Gaussian fits to the extended \Lya\
emission lines in  A1795 and A2597 are shown in Figures   
\ref{a1795tot}, \ref{a2597tot}. 
The results of the Gaussian fits are presented in Table \ref{tabemiss}. 
A fairly  conservative upper limit on the 
absorption EW in these spectra is 0.1A, or 25 km/s. We use the curve of
growth analysis in Laor (1997) to convert the EW to a column density. 
Laor gives convenient analytic approximations for the three regimes 
in the curve of growth (linear, saturated, and damped).
When the absorption becomes optically thick the absorption  EW $\simeq 
1.66*b$, and so the optically thin case applies
for EW=0.1A when $b>15$ km/s, and in that regime $N(HI) \simeq 1.8\times 10^{13}$
cm\mtwo. The transition to damped absorption occurs when $b<4$ km/s, 
which happens when $T<10^3$ K, and in that case 
$N(HI) \simeq 5.9\times 10^{16}$ cm\mtwo.
We suspect that such cold and kinematically quiet HI inside clusters is quite 
unlikely, and so the $b>15$ km/s HI column estimate is probably much 
more realistic.  Thus, we find an implied column density of $\lae  10^{13}$ 
cm\mtwo\ if the Doppler parameter $b>13$ km/s. 

Alternately, the covering factor of any high column density gas must 
be less than $\sim 25$\%. We note that in PKS 2322-123, the extended 
21 cm absorption is consistent with a covering factor as low as 
$6\times 10^{-3}$ (O'Dea, Baum \& Gallimore 1994).

\section{IMPLICATIONS}

Many mechanisms have been proposed to explain the lack of cooling gas
below temperatures of about 1 keV in ``cooling flow" clusters (e.g.,
Peterson \etal\ 2001, Fabian \etal\ 2001). Our results are relevant to
models which invoke differential absorption by cold gas to hide the 
missing low temperature X-ray gas. 
Fabian \etal\ estimate
that the required atomic hydrogen column densities are in the range of 
a few $\times 10^{21}$ to $10^{22}$ cm\mtwo. 
Such high columns would have been easily seen in
\Lya\ absorption since our upper limits are of order $\sim  10^{13}$
cm\mtwo.
The covering factors would need to be fairly high in order to reduce the 
emission from the cooling gas by observed factors of 5-10 (e.g., Peterson
\etal\ 2003). Fabian \etal\ note that the cold gas need not have a uniform
distribution in the ICM, but must preferentially cover the regions where
gas is cooling. Since our observations sample the lines of sight to the
emission line nebula (which are presumably the locations of the cooling gas) we
directly constrain this reduced covering factor model. 
Thus, we find that the ``missing'' cooling gas in cooling core
clusters is unlikely to be hidden by absorption.  

More generally,  we place limits on the column density of cool gas along our line of
sight to the bright central emission line nebulae. 
Thus, our results constrain the amount of cool gas embedded in the hot ICM,
i.e., outside the dense nebulae in the central $\sim 10-20$ kpc of the cluster
(e.g., Edge 2001; Jaffe, Bremer \& Baker 2005). 
If we assume that absorption feature 10 is one such system in the ICM, 
the column density of $2 \times 10^{14}$ cm\mtwo\ implies a total mass of
$\sim 40000$M$_\odot$ of cold gas in the hot ICM within a shell of radius  $\sim 20 - 100$ kpc.
If dense cold gas is injected into the ICM, e.g., by ram pressure stripping from galaxies
(e.g., Soker \etal\ 1991) or cooling on the $\sim 100$ kpc scale,  it can sink to
the center of the cluster on a time scale no shorter than  $\sim 100$ kpc/ 1000 km s\mone\ 
$\sim 10^8$ yr.  
This limits the rate at which cold gas can accumulate in the ICM on the 100 kpc
scales. 
The accumulation rate (i.e., the difference between the sources and sinks of cold gas)
is constrained to be no more than about $4 \times 10^{-4}$ M$_\odot$ yr\mone. 
This suggests (1) if cold clouds are not efficiently destroyed, injection of cold gas
into the hot ICM (by any mechanism, including cooling) on the 100 kpc scales is very low 
$\sim 10^{-4}$ M$_\odot$/yr; or (2)  any cold clouds on the 100 kpc scales are 
efficiently re-heated or destroyed (e.g., Loewenstein \& Fabian 1990); and (3) 
the dense nebulae in the cluster center are produced by gas which is deposited 
or which cools {\it in situ}.


\section{SUMMARY}

We present HST STIS long-slit spectroscopy of the \Lya\ line in A426,
A1795 and A2597. We detect ten \Lya\ absorption systems towards the
nucleus of NGC1275 with estimated column densities in the range
$N(HI) \sim 10^{12}$ to $10^{14}$ cm\mtwo. These systems could not have
been detected in the 21 cm line, but are easily detected in \Lya\
absorption. Most of the detected features are
located in the broad wings of the emission line and are beyond the
velocity range of the emission line filaments. 
The detected absorption systems are most consistent with 
associated nuclear absorption systems. Further observations of variability and
/or the metal lines are necessary to confirm this hypothesis. 

We do not detect the feature at 1237\AA\ reported by Johnstone \& Fabian
(1995) and interpreted as being due to Fermi-accelerated \Lya. 
If real, this feature is variable and would be consistent with an intrinsic
absorbing system. 

There is very little absorption at the systemic velocity of NGC1275 (feature
8 contains $N(HI) \sim 3\times 10^{12}$ cm\mtwo). This implies that the very
large column densities detected in the 21 cm line avoid the line of sight
to the nuclear \Lya\ emitting region and are likely detected against the
parsec scale radio jet. This atomic gas may be located in
a circumnuclear disk or torus. 

We detect two absorption systems (one at the systemic velocity)  towards the 
extended \Lya\ emission in A426. 

We do not detect \Lya\ absorption towards the extended diffuse 
\Lya\ emission in A1795 and A2597, with upper limits $N(HI) \lae 
 10^{13}$ cm\mtwo\ for optically thin absorbers with unity 
covering factor. Alternately, our data constrain the covering 
factor of any high column density gas ($N(HI) \gae 10^{15}$
cm\mtwo) to be less than 25\%. 

Our results suggest that it is unlikely that the ``missing" gas 
at temperatures below 1 keV in the cooling cores is due to 
absorption by large columns of absorbing gas with of order unity covering
factor. 
In addition, the low columns of gas on the $\sim 100$ kpc scales
in the ICM suggests that (1) the rate at which cold gas accumulates in the
ICM on these scales is very low, and (2) the dense nebulae in the central
$\sim 10$ kpc must have cooled or been deposited {\it in situ}. 


\acknowledgements

We are grateful to Andy Fabian, Jerry Kriss, and Rajib Ganguly
 for helpful discussions. We thank the anonymous referee for helpful 
comments. 
Support for program 8107 was provided by NASA through a grant from the
Space Telescope Science Institute, which is operate by the Association
of Universities for Research in Astronomy, Inc., under NASA contract
NAS 5-26555. This research made use of
(1) the NASA/IPAC Extragalactic Database
(NED) which is operated by the Jet Propulsion Laboratory, California
Institute of Technology, under contract with the National Aeronautics and
Space Administration; and (2)  NASA's Astrophysics Data System Abstract
Service.



\clearpage
\begin{deluxetable}{lrrr}
\tablewidth{0pt}
\tablecaption{Source Properties \label{tabsource}}
\tablehead{
\colhead{Parameter } &
\colhead{A~426/NGC1275 }  &
\colhead{A~1795 }  &
\colhead{A~2597}
}
\startdata
Second Name  & 3C84 &  4C~26.42  & PKS~B2322-123  \\
V Magnitude  & 12.5 &  14.2 &    15.8 \\
Galactic E(B-V) & 0.171     &  0.013 & 0.030 \\
Observed \Ha/\Hb\ & 4.77  & 3.2 & $4.2\pm 0.1$ \\ 
Redshift &  0.017559 &  0.06326  & 0.08220 \\
Scale (kpc/arcsec) & 0.33 & 1.12   & 1.42  \\
1.4 GHz Flux Density (Jy)  & 21.2 & 1.0 & 2.0 \\
Radio Power log$_{10}$P$_{ 1.4 GHz}$ (Watts Hz\mone) & 25.51 & 
25.31 & 25.84  \\
\enddata
\tablecomments{
We adopt a Hubble constant of $H_o = 75$ km s\mone\ Mpc\mone\ and
a deceleration parameter of $q_o = 0.0$. Balmer decrements are
from Kent \& Sargent (1979), Hu \etal\ (1985) and 
Voit \& Donahue (1997) respectively. 
 }
\end{deluxetable}

\begin{deluxetable}{llrrr}
\tablewidth{0pt}
\tablecaption{HST Spectroscopic Observations  \label{tabobs}}
\tablehead{
\colhead{Source } &
\colhead{Date  }  &
\colhead{Position Angle  }& 
\colhead{Exp. Time} &
\colhead{Central $\lambda$}
\\
\colhead{ } &
\colhead{ }  &
\colhead{degrees }&
\colhead{Secs. } &
\colhead{\AA  }
\\
\colhead{(1) } &
\colhead{(2) }  &
\colhead{ (3) }&
\colhead{ (4) } &
\colhead{ (5)  }
}
\startdata
A 426  & 05-12-2000 & 100     & 4900   & 1222 \\
A 1795 & 17-03-2001 & -158    & 4500   & 1272 \\
A 2597 & 18-11-2000 & 19      & 4300   & 1321 \\
\enddata
\tablecomments{
Observations were obtained  with the STIS FUV-MAMA and
the G140M grating through the $52 \times 0\farcs1$ slit under 
program 8107. TIME-TAG mode was used.  
Column (1) is the cluster name, (2) is the date of the observation, 
(3) is the slit position angle on the sky, (4) is the total exposure time,
and (5) is the central wavelength of the G140M grating. 
 }
\end{deluxetable}

\begin{deluxetable}{ll}
\tablewidth{0pt}
\tablecaption{Geocoronal Oxygen Lines \label{tabgeo}}
\tablehead{
\colhead{$\lambda$ } &
\colhead{Flux  }  
\\
\colhead{\AA } &
\colhead{ergs s\mone cm\mtwo arcsec\mtwo }
}
\startdata
1302.04  & 2.938E-12  \\
1304.73 & 2.974E-12  \\
1305.904 & 2.093E-12 \\
\enddata
\tablecomments{
Observed wavelength and integrated flux of the Geocoronal Oxygen lines detected in the G140M spectrum
of A2597.
The spectra were summed 10 arcsec along the slit. }
\end{deluxetable}

\begin{deluxetable}{llllrl}
\tablewidth{0pt}
\tablecaption{Absorption Lines Detected Towards NGC1275 \label{tablines}}
\tablehead{
\colhead{Feature } &
\colhead{EW } &
\colhead{$\lambda$  } &
\colhead{ v   }&
\colhead{ N(H) } &
\colhead{ b  } 
\\
\colhead{ } &
\colhead{\AA } &
\colhead{\AA } &
\colhead{  \kms }&
\colhead{  $10^{14}$ cm\mtwo }&
\colhead{  \kms }
\\
\colhead{ (1) } &
\colhead{ (2) }  &
\colhead{ (3) } &
\colhead{ (4) } &
\colhead{ (5) } &
\colhead{ (6) }
}
\startdata
1 &  0.5383 & 1223 & 1718 & 1.70 & 70 \\
2 &  0.1715 & 1225 & 2270 & 0.50 & 25 \\
3 &  0.6398 & 1226 & 2458 & 2.40 & 70 \\
4 &  0.1331 & 1231 & 3857 & 0.30 & 35 \\
5 &  0.3093 & 1232 & 4037 & 1.10 & 35 \\
6 &  0.3232 & 1233 & 4205 & 1.05 & 40 \\
7 &  0.2666 & 1233 & 4331 & 0.75 & 40 \\
8 &  0.1567E-01 & 1237 &  5264 & 0.03 & 15 \\
8$^a$ &  ...  & 1237   & 5264 & 0.80 & 80 \\
9 &  0.2817 & 1238 & 5585 & 0.67 & 60 \\
10 & 0.5348 & 1239 & 5794 & 1.90 & 60 \\
11$^{ab}$ & ... & 1237 & 5259 & ...  & ...  \\
\enddata
\tablecomments{
Properties of the detected absorption lines. 
Column (1) is the assigned number. Column (2) 
is the Equivalent Width, Column (3) is the observed
central wavelength of the line, Column (4) 
 is the heliocentric velocity
of the line. Column (5) is the estimated column density of 
atomic Hydrogen assuming a covering factor of unity. Column (6)
is the Doppler b factor ($b = \sigma 2^{1/2}$, and $\sigma$ is the Gaussian
dispersion). 
$^a$ Detected towards the extended emission. $^b$ Uncertain detection. 
 }
\end{deluxetable}

\begin{deluxetable}{rlccrrr}
\tablewidth{0pt}
\tablecaption{Candidate Identifications \label{tabgalaxy}}
\tablehead{
\colhead{Feature} &
\colhead{Galaxy  }  &
\colhead{Type   }&
\colhead{Mag  } &
\colhead{V$_{\rm line}$ - V$_{\rm ID}$ } &
\colhead{Offset  } &
\colhead{Offset }
\\
\colhead{ } &
\colhead{ }  &
\colhead{ }&
\colhead{ } &
\colhead{ \kms } &
\colhead{arcsec } &
\colhead{kpc  }
\\
\colhead{(1) } &
\colhead{(2) }  &
\colhead{(3) }&
\colhead{(4) } &
\colhead{(5) } &
\colhead{(6) } &
\colhead{(7)}
}
\startdata
4    & Per152    & E & 15 & -80 & 203 & 67 \\
4    & PGC012423 & E & 17 & -106& 238 & 79 \\
5    &    "      & E & 17 & 74  & 238 & 79 \\
9    & PGC012441 & E & 16 & 99  & 94  & 31 \\
10   & PGC012433 & E?& 18 & -27 & 92  & 31 \\
\enddata
\tablecomments{
Results from a NED search using V=1000 to 6000 \kms\ within
a box 8 arcmin across centered on NGC1275. There are two candidate
identifications for Feature 4. 
Column (1) The number of the absorption feature (see Table \ref{tablines}). 
Column (2) The name of the galaxy which is a candidate for 
identification with the absorption line. Column (3) The Hubble type
of the galaxy. Column (4) The magnitude of the candidate galaxy. 
Column (5) The offset in velocity between the absorption line and
the candidate galaxy. Column (6) The angular separation between the
candidate galaxy and the nucleus of NGC1275. Column (7) the offset in
kpc assuming the system is at the distance of the Perseus cluster.  
}
\end{deluxetable}

\begin{deluxetable}{lrcccc}
\tablewidth{0pt}
\tablecaption{Properties of the Extended \Lya\ Emission \label{tabemiss}}
\tablehead{
\colhead{Galaxy } &
\colhead{Slit Length  }  &
\colhead{Flux   }&
\colhead{Central $\lambda$ } &
\colhead{FWHM } &
\colhead{FWHM } 
\\
\colhead{ } &
\colhead{arcsec  }  &
\colhead{ergs s\mone cm\mtwo  }&
\colhead{ \AA } &
\colhead{ \AA } &
\colhead{ \kms }
}
\startdata
A426  & 5.8  & $2.4 \times 10^{-12}$  & 1236.67 & 3.2 & 780  \\
A1795 & 12.9 & $4.9\times 10^{-12}$  & 1291.80 & 2.7 & 630  \\
A2597 & 18.7 & $6.8\times 10^{-12}$  & 1315.79 & 3.7 & 844 \\
\enddata
\tablecomments{
The results of Gaussian fits to the spectra. 
The Slit Length is the distance along the slit that was summed to
produce the spectrum. }
\end{deluxetable}

\clearpage

\clearpage

\clearpage

\clearpage

\begin{references}

\reference{} Allen, S. W., Fabian, A. C., Johnstone, R. M., White, D. A., Daines,
S. J., Edge, A. C., \& Stewart, G. C. 1993, MNRAS, 262, 901

\reference{} Allen, S. W. \& Fabian, A. C. 1994, MNRAS, 269, 409

\reference{} Allen, S.~W. 1995, \mnras, 276, 947

\reference{} Allen, R. J., Knapen, J. H., Bohlin, R., \& Stecher, T. P.,
1997, \apj, 487, 171




\reference{} Bahcall, J. N., \& Ekers, R. D., 1969, ApJ, 157, 1055

\reference{} Bahcall, J. N., \etal, 1996, \apj, 457, 19





\reference{} Baum, S.~A., \& O'Dea, C. P., \mnras, 250, 737


\reference{} Binette, L., Joguet, B., \& Wang, J. C. L., 1998, \apj, 505,
634


\reference{} Blanton, E. L., Sarzain, C. L., \& McNamara, B. R., 2003, \apj, 585,
227

\reference{} B\"ohringer, H., \& Fabian, A. C., 1989, MNRAS, 237, 1147

\reference{} B\"ohringer, H., Voges, W., Fabian, A. C., Edge, A. C., 
\& Neumann, D. M., 1993,  MNRAS, 264, L25 

\reference{} B\"ohringer, H., Matsushita, K., Churazov, E., Ikebe,
Y., Chen, Y., 2002, A\&A,382, 804

\reference{} Boselli, A., Lequeux, J., \& Gavazzi, G., 2002, \aap, 384, 33 






\reference{} Burns, J. O., White, R. A., \& Haynes, M. P., 1981, AJ, 86, 1120




\reference{} Conselice, C. J., Gallagher, J. S., \& Wyse, R. F. G.,
2001, ApJ, 122, 2281

\reference{} Cowie, L. L. \& Binney, J. 1977 ApJ., 215, 723

\reference{} Crane, P. C., van der Hulst, J. M., \& Haschik, A. D. 1982,
in proceedings of IAU Symposium No. 97, Extragalactic Radio Sources,
eds. D. S. Heeschen \& C. M. Wade (Dordrecht, Reidel), 307

\reference{} Crawford, C.~S., Fabian, A.~C., Johnstone, R. M., \&
Crehan, D. A., 1987, \mnras, 224, 1007 



\reference{} Crenshaw, D. M., Kraemer, S. B., Boggess, A., Maran, S. P.,
Mushotzky, R. F., \& Wu, C.-C., 1999, ApJ, 516, 750






\reference{} Dwarakanath, K. S., van Gorkom, J. H., \& Owen, F. N. 1995,
ApJ, 442, L1


\reference{} Edge, A.~C., 2001, \mnras, 328, 762


\reference{} Ettori, S., Fabian, A. C., Allen, S. W., \& Johnstone, R. M.,
2002, \mnras, 331, 635

\reference{} Fabian, A. C., \& Nulsen, P. E. J., 1977, MNRAS, 180, 479

\reference{} Fabian, A.~C. 1994, \araa, 32, 277

\reference{} Fabian, A.~C., \etal\ 2000, \mnras, 318, L65 

\reference{} Fabian, A.~C.,Mushotzky, R. F., Nulsen, P. E. J., \& Peterson, J. R.,
2001, \mnras, 321, L20

\reference{} Fabian, A.~C., Allen, S. W., Crawford, C. S., Johnstone, R. M., 
Morris, R. G., Sanders, J. S., \& Schmidt, R. W., 2002, \mnras, 332, L50 


















\reference{}  Jaffe, W., 1990, A\&A, 240, 254

\reference{}  Jaffe, W. 1991, A\&A, 250, 67

\reference{}  Jaffe, W. 1992, In proceedings of the NATO ASI ``Clusters and
Superclusters of Galaxies," ed. A. C. Fabian, (Kluwer, Dordrecht),  109

\reference{}  Jaffe, W., Bremer, M. N., \& Baker, K., 2005, astro-ph/0504413


\reference{} Johnstone, R.~M., \& Fabian, A.~C., 1995, \mnras, 273, 625

\reference{} Kaastra, J. S.,. Ferrigno, C., Tamura, T., Paerels, F. B. S.,
Peterson, J. R., \& Mittaz, J. P. D., 2001, \aap, 365, 99L

\reference{}  Kaastra, J. S., \etal, 2004, \aap, 413, 415


\reference{} Kimble, R. A., \etal, 1998, ApJ, 492, L83

\reference{} Klein, R., McKee, C., \& Colella, P., 1994, \apj, 420, 213

\reference{} Koekemoer, A.~M., O'Dea, C.~P., Baum, S.~A., Sarazin, C.~L., 
Owen, F. N., \& Ledlow, M. J., 1998, \apj, 508, 608 


\reference{} Kriss, G. A., \etal, 2000, ApJ, 538, L17


\reference{} Laor, A., 1997, \apjl, 483, L103


\reference{} Levinson, A., Laor, A., \& Vermeulen, R. C., 1995,
ApJ, 448, 589

\reference{} Loewenstein, M. \& Fabian, A. C. 1990, MNRAS, 242, 120


\reference{} Mathews, W. G. \& Bregman, J. N. 1978, ApJ, 224, 308



\reference{} McNamara, B. R., Bregman, J., N. \& O'Connell, R. W. 1990, 
ApJ, 360, 20



\reference{} McNamara, B.~R., O'Connell, R.~W., \& Sarazin, C. L., 1996,
AJ, 112, 91 




\reference{} Neufeld, D. A., \& McKee, C. F., 1988, \apjl, 331, L87 



\reference{}  Miller, E. D., Bregman, J. N., \& Knezek, P. M., 2001, \apj,
569, 134




 \reference{} Mushotzky, R. F. 1992, in proceedings of the NATO ASI ``Clusters and
Superclusters of Galaxies," ed. A. C. Fabian (Kluwer, Dordrecht), 91





\reference{} O'Dea, C. P., Baum, S. A., \& Gallimore, J. F., 1994, ApJ, 436,
669

\reference{} O'Dea, C. P., Gallimore, J. F., \& Baum, S. A., 1995, AJ, 109, 26



\reference{} O'Dea, C. P., Payne, H., \& Kocevski, D., 1998, AJ, 116, 623

\reference{} O'Dea, C. P., \etal, 2003, PASA, 20, 88

\reference{} O'Dea, C. P., Baum, S. A., Mack, J., Koekemoer, A., \&
Laor, A., 2004, \apj, 612, 131 


\reference{} Peterson, J. R., \etal, 2001, \aap, 365, 104L

\reference{} Peterson, J. R., \etal, 2003, \apj, 590, 207



\reference{} Rauch, M. 1998, ARA\&Ap, 36, 267

\reference{} Rybicki, G. \&  Lightman, A. P., 1979, Radiative Processes in Astrophysics,  
(Wiley-Interscience, New York)





\reference{} Sarazin, C. L. 1988, X-ray Emission from Clusters of Galaxies,
(Cambridge, Cambridge University Press)




\reference{} Sijbring, D., 1993, PhD dissertation, Univ. of Groningen



\reference{} Soker, N., Bregman, J. N., \& Sarazin, C. L., 1991, \apj,
368, 341

\reference{} Soker, N., Blanton, E. L., \& Sarazin, C. L., 2002, \apj,
573, 533

\reference{} Sparks, W.~B., Macchetto, F. D., \& Golombek, D., 1989,
\apj, 345, 153


\reference{} Sparks, W.~B.,  1992, ApJ, 399, 66


\reference{} Tamura, T., \etal\ 2001, \aap, 365, L87


\reference{} Taylor, G. B., 1996, ApJ, 470, 394

\reference{} Taylor, G. B., O'Dea, C. P., Peck, A. B., \& Koekemoer,
A. M., 1999, ApJ, 512, 27L


\reference{} Tucker, W., \& David, L. P., 1997, \apj, 484, 602

\reference{}  Valentijn, E. A., \& Giovanelli, R. 1982, A\&A, 114, 208



\reference{} Vermeulen, R. C., Readhead, A. C. S., \& Backer, D. C.,
1994, ApJ, 430, L41



\reference{} Walker, R. C., Romney, J. D., \& Benson, J. M., 1994,
ApJ, 430, L45

\reference{} Walker, R. C., Dhawan, V., Romney, J. D., Kellermann, K.
I., \& Vermeulen, R. C., 2000, ApJ, 530, 233

\reference{} Weymann, R. J., Morris, S. L., Gray, M. E., \& Hutchings,
J. B., 1997, ApJ, 483, 717

\reference{} White, D. A., Fabian, A., C., Johnstone, R. M., Mushotzky, R. F.,
\& Arnaud, K. A. 1991, MNRAS, 252, 72


\end{references}
\end{document}